\documentclass[aps,prl,twocolumn,superscriptaddress]{revtex4-1}

\usepackage{graphicx}
\usepackage{dcolumn}
\usepackage{bm}
\usepackage{hyperref}
\usepackage{todonotes}
\usepackage{color}

\begin{document}

\title{Ultrafast energy and momentum resolved dynamics of magnetic correlations in photo-doped Mott insulator Sr$_2$IrO$_4$}
\author{M. P. M. Dean}
\email{mdean@bnl.gov; Contributed equally to this work}
\affiliation{Department of Condensed Matter Physics and Materials Science, Brookhaven National Laboratory, Upton, New York 11973, USA}

\author{Yue Cao}
\email{ycao@bnl.gov; Contributed equally to this work}
\affiliation{Department of Condensed Matter Physics and Materials Science, Brookhaven National Laboratory, Upton, New York 11973, USA}

\author{X. Liu}
\email{xliu@iphy.ac.cn} 
\affiliation{Beijing National Laboratory for Condensed Matter Physics and Institute of Physics, Chinese Academy of Sciences, Beijing 100190, China}
\affiliation{Collaborative Innovation Center of Quantum Matter, Beijing, China}

\author{S. Wall}
\affiliation{ICFO-Institut de Ci{\`e}ncies Fot{\`o}niques, The Barcelona Institute of Science and Technology, 08860 Castelldefels (Barcelona), Spain}

\author{D. Zhu}
\affiliation{Linac Coherent Light Source, SLAC National
Accelerator Laboratory, Menlo Park, California, USA}

\author{R. Mankowsky}
\affiliation{Max Planck Institute for the Structure and Dynamics of Matter, Hamburg, Germany}
\affiliation{Center for Free Electron Laser Science, Hamburg,Germany}

\author{V. Thampy}
\affiliation{Department of Condensed Matter Physics and Materials Science, Brookhaven National Laboratory, Upton, New York 11973, USA}

\author{X. M. Chen}
\affiliation{Department of Condensed Matter Physics and Materials Science, Brookhaven National Laboratory, Upton, New York 11973, USA}

\author{J. G. Vale}
\affiliation{London Centre for Nanotechnology and Department of Physics and Astronomy, University College London, London WC1E 6BT, UK}

\author{D. Casa}
\affiliation{Advanced Photon Source, Argonne National Laboratory, Argonne, Illinois 60439, USA}

\author{Jungho Kim}
\affiliation{Advanced Photon Source, Argonne National Laboratory, Argonne, Illinois 60439, USA}

\author{A. H. Said}
\affiliation{Advanced Photon Source, Argonne National Laboratory, Argonne, Illinois 60439, USA}

\author{P. Juhas}
\affiliation{Department of Condensed Matter Physics and Materials Science, Brookhaven National Laboratory, Upton, New York 11973, USA}

\author{R. Alonso-Mori}
\affiliation{Linac Coherent Light Source, SLAC National
Accelerator Laboratory, Menlo Park, California, USA}

\author{J. M. Glownia}
\affiliation{Linac Coherent Light Source, SLAC National
Accelerator Laboratory, Menlo Park, California, USA}

\author{A. Robert}
\affiliation{Linac Coherent Light Source, SLAC National
Accelerator Laboratory, Menlo Park, California, USA}

\author{J. Robinson}
\affiliation{Linac Coherent Light Source, SLAC National
Accelerator Laboratory, Menlo Park, California, USA}

\author{M. Sikorski}
\affiliation{Linac Coherent Light Source, SLAC National
Accelerator Laboratory, Menlo Park, California, USA}

\author{S. Song}
\affiliation{Linac Coherent Light Source, SLAC National
Accelerator Laboratory, Menlo Park, California, USA}

\author{M. Kozina}
\affiliation{Linac Coherent Light Source, SLAC National
Accelerator Laboratory, Menlo Park, California, USA}

\author{H. Lemke}
\affiliation{Linac Coherent Light Source, SLAC National
Accelerator Laboratory, Menlo Park, California, USA}

\author{L. Patthey}
\affiliation{SwissFEL, Paul Scherrer Institut, CH-5232 Villigen PSI, Switzerland}

\author{S. Owada}
\affiliation{RIKEN SPring-8 Center, Sayo, Hyogo 679-5148, Japan}

\author{T. Katayama}
\affiliation{Japan Synchrotron Radiation Institute, 1-1-1 Kouto, Sayo-cho, Sayo-gun, Hyogo 679-5198, Japan}

\author{M. Yabashi}
\affiliation{RIKEN SPring-8 Center, Sayo, Hyogo 679-5148, Japan}

\author{Yoshikazu Tanaka}
\affiliation{RIKEN SPring-8 Center, Sayo, Hyogo 679-5148, Japan}

\author{T. Togashi}
\affiliation{Japan Synchrotron Radiation Institute, 1-1-1 Kouto, Sayo-cho, Sayo-gun, Hyogo 679-5198, Japan}

\author{Jian Liu}
\affiliation{Department of Physics \& Astronomy, University of Tennessee, Knoxville, TN 37996, USA}
\author{C. Rayan Serrao}
\affiliation{Department of Electrical Engineering and Computer Sciences, University of California, Berkeley, California 94720, USA}

\author{B. J. Kim}
\affiliation{Max Planck Institute for Solid State Research, D-70569 Stuttgart, Germany}

\author{L. Huber}
\affiliation{Institute for Quantum Electronics, ETH Zurich, CH-8093 Zurich, Switzerland}

\author{C.-L. Chang}
\affiliation{Zernike Institute for Advanced Materials, University of Groningen,
Groningen, NL 9747AG, the Netherlands}

\author{D. F. McMorrow}
\affiliation{London Centre for Nanotechnology and Department of Physics and Astronomy, University College London, London WC1E 6BT, UK}

\author{M. F{\"o}rst}
\affiliation{Max Planck Institute for the Structure and Dynamics of Matter, Hamburg, Germany}
\affiliation{Center for Free Electron Laser Science, Hamburg,Germany}

\author{J. P. Hill}
\affiliation{Department of Condensed Matter Physics and Materials Science, Brookhaven National Laboratory, Upton, New York 11973, USA}

\def\mathbi#1{\ensuremath{\textbf{\em #1}}}
\def\Q{\ensuremath{\mathbi{Q}}}
\def\SIO{Sr$_2$IrO$_4$}
\newcommand{\angstrom}{\mbox{\normalfont\AA}}
\date{\today}
%
%
%

%
\maketitle

\textbf{Measuring how the magnetic correlations throughout the Brillouin zone evolve in a Mott insulator as charges are introduced dramatically improved our understanding of the pseudogap, non-Fermi liquids and high $T_C$ superconductivity \cite{Scalapino2012, Kim2014, Cao2014, delaTorre2015}. Recently, photoexcitation has been used to induce similarly exotic states transiently \cite{Fausti2001, Zhang2014, Aoki2014}. However, understanding how these states emerge has been limited because of a lack of available probes of magnetic correlations in the time domain, which hinders further investigation of how light can be used to control the properties of solids. Here we  implement magnetic resonant inelastic X-ray scattering at a free electron laser, and directly determine the magnetization dynamics after photo-doping the Mott insulator \SIO{}. We find that the non-equilibrium state 2~ps after the excitation has strongly suppressed long-range magnetic order, but hosts photo-carriers that induce strong, non-thermal magnetic correlations. The magnetism recovers its two-dimensional (2D) in-plane N{\'e}el correlations on a timescale of a few ps, while the three-dimensional (3D) long-range magnetic order restores over a far longer, fluence-dependent timescale of a few hundred ps. The dramatic difference in these two timescales, implies that characterizing the dimensionality of magnetic correlations will be vital in our efforts to understand ultrafast magnetic dynamics.}

\begin{figure*}
\includegraphics[width=6.in]{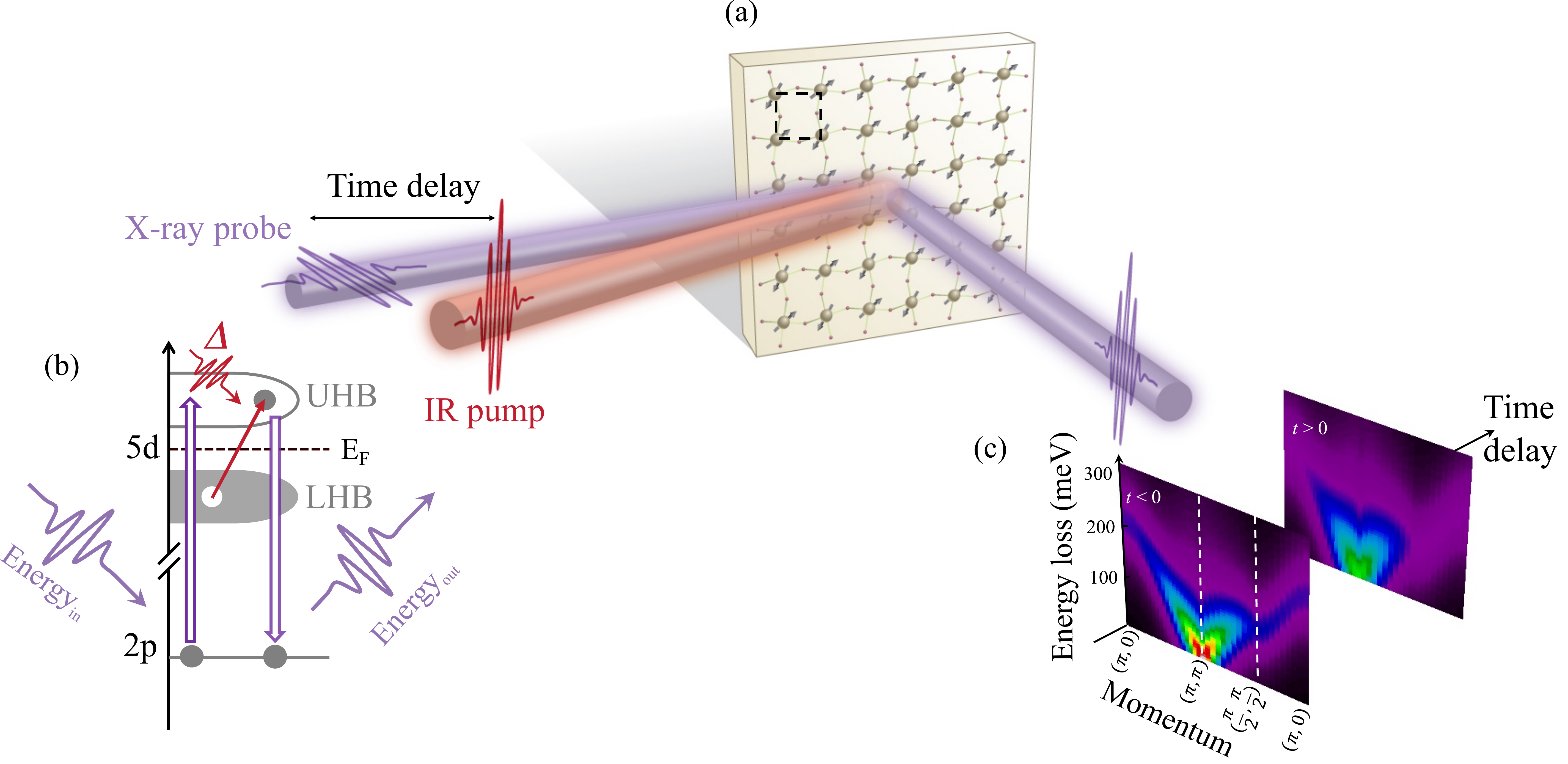} %
\caption{\textbf{Experimental configuration.} \textbf{a}, The scattering setup. The vertically polarized pump pulse (shown in red) is incident on the $ab$-face of \SIO{}. X-ray pulses from a free electron laser (shown in purple) probe the resulting transient state. X-rays that are scattered close to 90$^{\circ}$ are either directly measured, to access the magnetic Bragg peak that probes the presence or absence of 3D magnetic order, or energy analyzed to access the inelastic spectrum that is particularly sensitive to the 2D magnetic correlations. The basic in-plane structural unit of \SIO{} is outlined with a dotted black line. \textbf{b}, An illustration of the pump and probe processes. The 620~meV (2~$\mathrm{\mu m}$) pump beam (in red) photo-dopes the sample exciting an electron from the lower Hubbard band (LHB) to the upper Hubbard band (UHB). Horizontally polarized 11.215~keV X-ray pulses from a free electron laser (shown in purple) probe the resulting transient state. The incident X-ray pulses excite a Ir $2p$ core electron into the $5d$ valence band, in order to couple to the spin degree of freedom. The resulting emitted photon encodes the magnetic and orbital configuration of the transient state  \cite{Ament2011}. \textbf{c}, Illustration of the detection of x-rays as a function of energy loss, momentum transfer and time delay, encoding the time dependent magnetic correlations in the transient state. The RIXS planes plot simple spin wave calculations based on an increased thermal population of magnon after the pulse.}
\label{Fig1}
\end{figure*}

In the layered perovskite \SIO{}, multiple interactions conspire to determine its electronic configuration. Strong spin-obit coupling splits the Ir 5d states to form a narrow electronic band that can be further split by the modest on-site Coulomb repulsion to generate an antiferromagnetic Mott insulating state with close structural, and electronic analogies to the superconducting cuprates \cite{Kim2009, Kim2012, Kim2014, Cao2014, delaTorre2015}. It has been well established that when a perturbation melts magnetic order in a Mott insulator, the resulting new state frequently exhibits unusual properties \cite{Lee2006}. For example, surface-doping and Rh-Ir substitution in \SIO{} have generated novel Fermi-arc and pseudogap behavior \cite{Kim2014, Cao2014, delaTorre2015} and some have argued that doped \SIO{} might host high temperature superconductivity \cite{Wang2011, Yan2015}. In both cases, magnetic correlations were argued to play a critical role in the formation of these states. Photo-doping a Mott insulator using ultrafast lasers provides an alternative route to create transient versions of these exotic states, with the advantage that the resulting states are tunable and reversible. To date, however, the appropriate tools have been lacking for probing the momentum \emph{and} energy dependence of the electronic and magnetic correlations characterizing these ultrafast transient states.

Figure \ref{Fig1} illustrates our experimental approach. \SIO{} was cooled to 110~K, well below its N{\'e}el ordering temperature of 240~K \cite{Cao1998}. Pump laser pulses with an energy of 620~meV (2~$\mu$m) drive carriers from the lower Hubbard band to the upper Hubbard band \cite{Moon2006}. The transient magnetic response to this pump was characterized using a free electron laser. X-ray photons were tuned to the Ir $L_3$ resonance in order to couple to the spin degree of freedom via the resonant magnetic x-ray scattering mechanism and photons scattered around $90^{\circ}$ were measured as a function of momentum transfer, $\mathbi{Q}$, energy loss, $E$, and time delay, $t$.

\begin{figure*}
\includegraphics[width=7in]{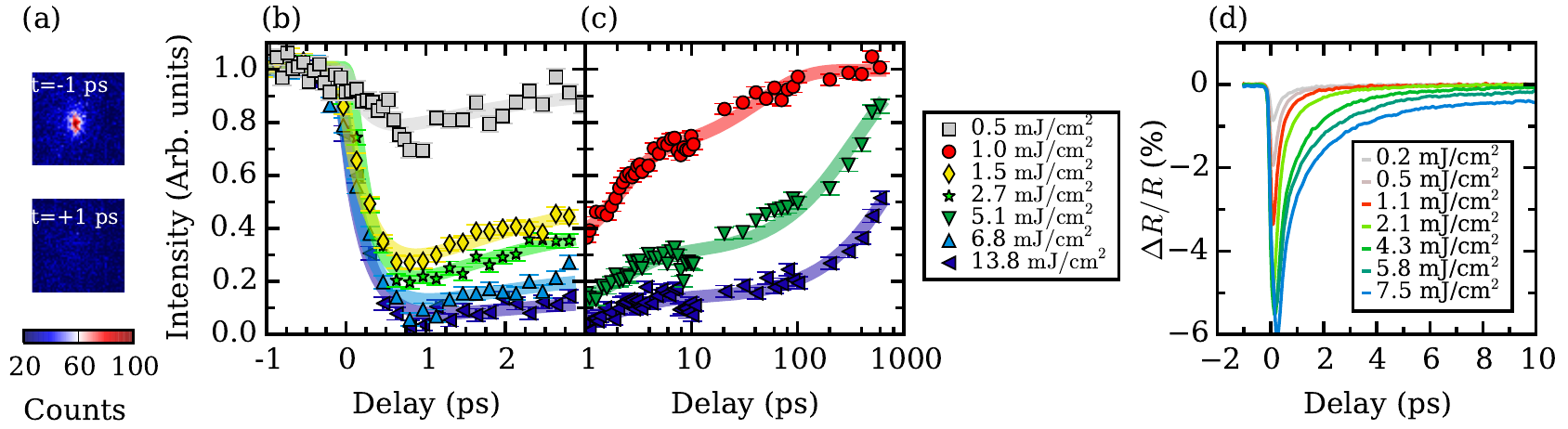} %
\caption{\textbf{Destruction and recovery of charge and 3D magnetic order in \SIO{}.} 
\textbf{a} Intensity of the $(-3,-2,28)$ magnetic Bragg peak 1~ps before (top panel) and 1~ps after (bottom panel)  excitation at 6.8~mJ/cm$^2$. \textbf{b} and \textbf{c} plot the intensity of the magnetic Bragg peak as a function of probe delay focusing on the short and long timescales respectively. The lines show the result of fitting a model, which incorporates one decay timescale and two recovery timescales. \textbf{d}, Relative change in the  800~nm optical reflectivity of \SIO{} after excitation with a 620~meV pump at different fluences. All data is taken at 110~K.}
\label{Fig2}
\end{figure*}

Figure \ref{Fig2}a,b plots the time and fluence dependence of the $(-3,-2,28)$ magnetic Bragg peak intensity in \SIO{}, which is sensitive to the presence of 3D magnetic order. This intensity is measured by an area detector without energy-analyzing the scattered photons. We find that fluences of $\gtrsim 5$~mJ/cm$^2$ destroy the 3D magnetic order based on the criterion of having $\lesssim 10$\% remnant intensity in the magnetic Bragg peak. This fluence corresponds to exciting a substantial fraction of all the lattice sites within the illuminated volume. Indeed, comparable fluences were also required to destroy long-range magnetic order in other strongly correlated materials including manganites \cite{Ehrke2011, Zhou2012} and nickelates \cite{Chuang2013, Caviglia2013, lee2012phase}. 

In order to characterize the charge response to the 620~meV (2~$\mu$m) pump excitation, we measured optical reflectivity at 1.55~eV (800~nm) in Fig.~\ref{Fig2}d. The photo-carrier recombination is dominated by processes in the ps or sub ps regime, far faster than the recovery of 3D magnetic order, suggesting that the charge and magnetic recovery processes are largely independent of one-another. 

A detailed understanding of ultrafast magnetic dynamics, beyond the presence or absence of 3D magnetic order, is severely hampered by limited experimental information regarding the short range transient magnetic correlations. Other existing techniques such as X-ray magnetic dichroism \cite{Boeglin2010}, Faraday rotation \cite{Kampfrath2011} and the magneto-optical Kerr effect \cite{Malinowski2008} capture only 3D magnetic order. This letter breaks new ground by energy analyzing the scattered X-rays i.e.\ by performing the first ever time resolved (tr) magnetic Resonant Inelastic X-ray Scattering (RIXS) experiment. RIXS probes the magnetic quasiparticle spectrum itself \cite{Ament2011, Dean2015}. This is a fundamental expression of the nature of the correlated electron state -- as it is the spatial and temporal Fourier transform of the spin-spin correlation function and it encodes the interactions present in the magnetic Hamiltonian. In the present case of the $5d$ valence electron compound \SIO{}, the relevant X-ray $L$-edge is in the hard X-ray regime allowing full access to reciprocal space. Such $\mathbi{Q}$-space resolution is not available in the complementary technique of time resolved two-magnon Raman scattering, due to the fact that visible photons carry negligible momentum \cite{Batignani2015}. 

\begin{figure*}
\includegraphics{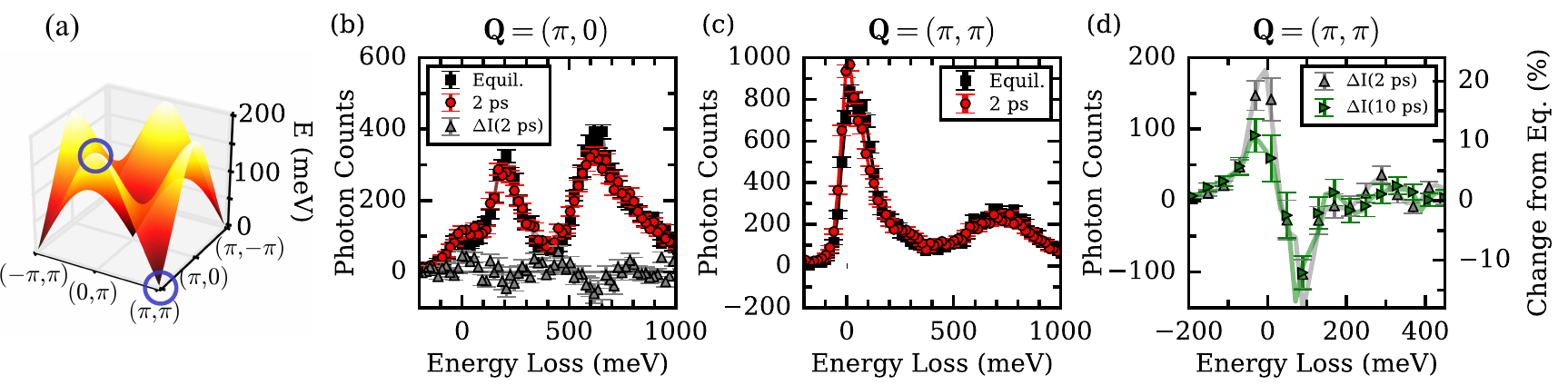} %
\caption{\textbf{2D magnetic correlations before and after photo-excitation.} \textbf{a}, Equilibrium state magnetic dispersions of \SIO{} based on a spin wave fit to measurements in Ref.~\cite{Kim2012}. The $\mathbi{Q}$-vectors studied are outlined in blue. \textbf{b},\textbf{c}  tr-RIXS  spectra showing magnetic excitations (0-200~meV) and orbital excitations ($\sim600$~meV) in the equilibrium state 50~ps before photo-excitation (labeled Equil.) and 2~ps after photo-excitation at 6~mJ/cm$^2$.  \textbf{b} Shows that high-energy nearest neighbor 2D magnetic correlations, as probed at $\mathbi{Q}=(\pi,0)$, have completely recovered 2~ps after the pump. \textbf{c}, Plots the relatively low energy magnetic fluctuations at $\mathbi{Q}=(\pi,\pi)$ that  arise from a small disturbance of the N\'{e}el order.  \textbf{d}, Difference spectra between the equilibrium state and the 2~ps transient state (from panel \textbf{c}) and between the GS and 10~ps . This shows a depletion of approximately 20\% of the magnetic spectral weight around $\sim 100$~meV and additional spectral intensity appearing at very low energy.}
\label{Fig3}
\end{figure*}

Figure \ref{Fig3} plots the RIXS energy loss spectra measured in \SIO{} after photo-excitation at 6~mJ/cm$^2$ , as compared to the unperturbed state 50~ps before excitation. The chosen pump fluence corresponds to what was required to destroy 3D magnetic order, as seen in Fig.~\ref{Fig2}a,b. The RIXS spectra show two dominant features, identified as magnon and orbital excitations \cite{Kim2012, Ishii2011, Kim2014excitonic}, which we address in turn. 

Orbital excitations appear around 600~meV, and correspond to exciting an electron from the $J_{\text{eff}}=\frac{1}{2}$  ground state orbital to the $J_{\text{eff}}=\frac{3}{2}$ state \cite{Ishii2011, Kim2014excitonic}. The intensity of the orbital excitation is proportional to the $J_{\text{eff}}=\frac{1}{2} \rightarrow \frac{3}{2}$ transition cross-section, and thus directly reflects the electron population in these orbitals. This excitation is different from the pump excitation at 620~meV in that the RIXS process involves two dipole transitions so the process is allowed on a single site \cite{Ament2011}. Given that a very similar amplitude of orbital excitations are seen before and after excitation, we infer that the vast majority of the photo-excited carriers have decayed out of the $J_{\text{eff}}=\frac{3}{2}$ orbital even in this relatively short time window. This sets an upper limit of 2~ps on the lifetime of electron-hole pairs, or doublons, living in the upper and lower Hubbard bands respectively in \SIO{}. We know from our optical reflectivity measurements in Fig.~\ref{Fig2}d that an appreciable population of photo-excited carriers persists at 6~mJ/cm$^2$ and 2~ps delay. These could exist either via a spectral weight redistribution of the Hubbard bands or as in-gap states \cite{Okamoto2010}.

The equilibrium excitation spectrum of undoped \SIO{} can be accurately modeled in terms of magnons, or spin waves, that arise from Heisenberg spin-spin exchange interactions \cite{Kim2012}. Figure~\ref{Fig3}a plots the magnetic dispersion from low energies at $(\pi,\pi)$ to $\sim 200$~meV at $(\pi,0)$. Corresponding experimental RIXS spectra are plotted in Fig.~\ref{Fig3}b,c with data shown both in the equilibrium (50~ps before the pump) and in the transient state after the pump. Despite the almost-complete destruction of the magnetic Bragg peak in the transient state, magnons are still observed at both $\mathbi{Q}$ points. Due to the relatively weak $c$-axis exchange interaction in \SIO{}, the intensity, energy-scale and dispersion of these magnons is most sensitive to the 2D N{\'e}el correlations between neighboring spins \cite{Kim2012}. This indicates that 2D correlations largely retain their N{\'e}el-like nature in the transient state 2~ps after the pump. The fact that the magnetic Bragg peak, as shown in Fig.~\ref{Fig2}a,b, is very strongly attenuated in the transient state is likely to be due predominantly to the destruction of inter-plane correlations along the $c$-axis. Looking at the RIXS difference spectra in detail, we see that the magnon at $(\pi,0)$ is identical before and after the pump. At $(\pi,\pi)$, however, there is an appreciable change. This indicates that the high energy $\sim 200$~meV correlations at $(\pi,0)$ are more robust than the lower energy spin wave at $(\pi,\pi)$ that arises from a smaller disturbance of the N{\'e}el order. One interpretation of this observation is that the higher energy magnons recover to their equilibrium configuration in much less than 2~ps, which could be due to the fact that a higher energy excitation can decay into lower energy multi-particle excitations in a larger number of different ways than can the lower energy magnons.

We now consider the magnetic excitation spectrum around $(\pi,\pi)$ in more detail. Due to the finite ($\pm 0.5$~\AA$^{-1}$) momentum resolution of our spectrometer, the observed spectrum is the sum of the very low energy magnons from precisely $\mathbi{Q}=(\pi,\pi)$, and of slightly higher energy magnons from closely neighboring $\mathbi{Q}$ values. This leads to the asymmetric peak in Fig.~\ref{Fig3}c, which is further plotted as transient state difference spectra in Fig.~\ref{Fig3}d. We see that $\sim 20$\% of the spectral intensity around $\sim 100$~meV has been depleted and additional very low energy spectral intensity appears, which recovers on a ps timescale. Thermal heating effects on 2D quantum antiferromagnets have been studied extensively and result in a uniform relative broadening of the magnons across the Brillouin zone \cite{Manousakis1991, Ronnow2001}. Such broadening is not observed here, excluding a purely thermal explanation of our results. Given that we have established the presence of residual photo-excited carriers in the transient state, we suggest that these carriers are directly responsible for damping the magnetic correlations around $\sim 100$~meV and causing an apparent redistribution of the magnetic spectral weight to lower energy. 

Having clarified the 2D correlations in the transient state, we reassess the behavior of the 3D magnetic order presented in Fig.~\ref{Fig2}. Even in the few ps regime (Fig.~\ref{Fig2}b), a small amount of magnetic recovery is evident. However, full recovery takes somewhere between 100 to over 1000~ps (Fig.~\ref{Fig2}c). We found that a minimal model for the magnetic intensity as a function of time, $I(t)$, required one decay timescale $\tau_{\text{decay}}$ and two recovery timescales, which for reasons that we will explain later, are labeled $\tau_{\text{2D}}$ and $\tau_{\text{3D}}$, where $\tau_{\text{2D}} < \tau_{\text{3D}}$

\begin{eqnarray}
I(t) & = & I_0 \Big( \exp(-t/\tau_{\text{decay}}) + C \big[1 - \exp(-t/\tau_{\text{2D}})\big]  \nonumber \\
 & + & \left(1-C\right) \big[1 - \exp(-t/\tau_{\text{3D}}) \big]  \Big).
\label{equation}
\end{eqnarray}

This model was fit to the magnetic Bragg peak intensity data in Fig.~\ref{Fig2}b,c. In a similar way, we fitted the recovery of the optical reflectivity, which also required two charge timescales denoted $T_{\text{fast}}$ and $T_{\text{slow}}$.

\begin{figure*}
\includegraphics{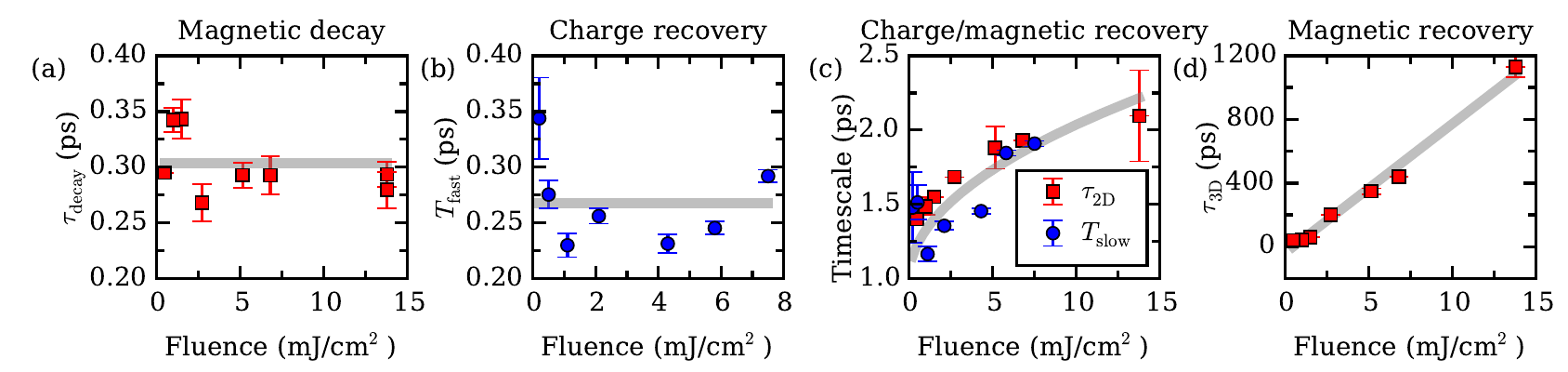} %
\caption{\textbf{Fluence dependence of the magnetic and charge dynamics timescales.}  \textbf{a}, the magnetic decay timescale $\tau_{\text{decay}}$, showing that the magnetic decay happens in $\leq 0.3$~ps, faster than the time resolution of the experiment. \textbf{b}, the fast charge recovery timescale $T_{\text{fast}}$. \textbf{c}, the timescales for the slow component of the charge recovery $T_{\text{fast}}$ and the faster component of the magnetic recovery. From our RIXS data we know that the 2D magnetic correlations recover on a ps timescale, so due to the similar timescale of this fitting parameter it is labelled $T_{\text{2D}}$. These show strikingly similar magnitudes and fluence dependencies. \textbf{d}, $\tau_{\text{3D}}$ the slow magnetic recovery, which is assigned to the restoration of 3D magnetic order.  Grey lines are guides to the eye and error bars are the uncertainty from the least square fitting algorithm.}
\label{Fig4}
\end{figure*}

Figure \ref{Fig4} summarizes the magnetic and charge dynamics in \SIO{} after laser-excitation creates a large population of doublons. 3D magnetic order decays in $0.30 \pm 0.03 $~ps approximately independent of fluence (Fig.~\ref{Fig4}a), which is roughly equal to the jitter-limited time resolution of the experiment. This sets an upper limit on the timescale for the destruction of magnetic order in this system. The fast component of the charge recovery (Fig.~\ref{Fig4}b) is of a similar magnitude $0.27 \pm 0.04$~ps. Panel c plots the faster magnetic recovery timescale which is $\sim 1.7$~ps and increases slowly with fluence. As discussed above, the 2D in-plane magnetic correlations recover on the ps timescale, so the lack of 3D magnetic order can be primarily attributed to the lack of coherence between the IrO$_2$ planes. Or, equivalently, after a few ps there is a large population of low-energy long-wavelength magnons that, on average, preserve the 2D N{\' e}el correlations. Consequently, we attribute the fast timescale to 2D in-plane correlations and label it $\tau_{\text{2D}}$. On the hundreds of ps timescale the 2D correlation are largely restored and the recovery is dominated by the restoration of 3D inter-plane correlations, leading us to assign the slower timescale to $\tau_{\text{3D}}$. The dramatic difference in these timescales reflects the strong anisotropy in the magnetic interactions. \SIO{} has very strong in-plane magnetic exchange $J_{\parallel}= 60$~meV and very weak inter-plane magnetic exchange, which some researchers have estimated to be as low as $J_{\perp} \approx 1$~$\mathrm{\mu}$eV \cite{Kim2012, Fujiyama2012, Kim2014excitonic, Vale2015}. 
$\tau_{\text{3D}}$ is also seen to be linear with fluence (Fig.~\ref{Fig4}d), increasing up to 1130~ps at 13.8~mJ/cm$^2$. This implies that the recovery of the long-range magnetic order relies crucially on the dissipation of energy in the material and is set by the timescale upon which the low-energy long-wavelength magnons can dissipate their energy into the lattice degree of freedom, whereas $\tau_{\text{2D}}$ has far weaker fluence dependence and seems to be related to a material property.

Time-resolved resonant X-ray scattering has provided a new window on the transient magnetic state in photo-doped \SIO{}. The 2D magnetic correlations we observe are non-thermal in nature and recover on a ps timescale denoted $\tau_{\text{2D}}$. A striking similarity between $\tau_{\text{2D}}$ and the slower charge recovery timescale $T_{\text{slow}}$ is seen in Fig.~\ref{Fig4}c. This may reflect the similar energy scale of the in-plane electronic hopping parameter, $t_\parallel$, and the magnetic exchange, $J_\parallel$, which are fundamentally linked in strongly correlated materials such as these via $J_\parallel \sim t^2_\parallel / U$ where $U$ is the Coulomb repulsion. The behavior of the long-range magnetic order, in contrast, depends on secondary processes, such as inter-plane magnetic coupling and the dissipation of the energy from the spins into other degrees of freedom.  

This work shows that direct measurements of the 2D magnetic correlations are consequently crucial for a full understanding of magnetic dynamics in strongly correlated materials. With the continued improvement of free electron lasers, tr-RIXS is set to play a crucial role in understanding how magnetic correlations dictate the properties of doped Mott insulators and how they can be effectively manipulated by light.

\section*{Methods\label{Methods}}
\noindent\textbf{Samples}
The magnetic Bragg peak measurements were performed on 200~nm epitaxial films of \SIO{}, in order to match the volume of \SIO{} to the penetration depth of the pump, as the X-ray penetration depth is  longer than the pump. The disappearance of the magnetic Bragg peak in Fi.~\ref{Fig2}a,b confirms that the whole probed volume is excited. The film was deposited on SrTiO$_3$ using pulsed laser deposition as described in the supplementary information and Ref.~\cite{Serrao2013}. For RIXS, $\sim1^\circ$ grazing incidence X-rays were used to limit X-ray penetration depths to 80~nm on a bulk \SIO{} crystal. Both samples have a $c$-axis surface normal. Reciprocal lattice notations are defined using the full unit cell with lattice constants $a=b=5.51$~\AA{} and $c=25.7$~\AA{}. The high symmetry points in the in-plane Brillouin zone are defined in the reduced structural zone (which ignores the rotation of the IrO$_6$ octahedra) as in Ref.~\cite{Kim2012}. The zone center, denoted $(\pi, \pi)$ and the zone boundary denoted $(\pi,0)$ correspond to $(1,0,L)$ and $(0.5, 0.5, L)$ respectively in the reciprocal lattice notation. In both experiments the sample was cooled to about 110~K with a stream of nitrogen gas, well below the N\'{e}el temperature of 240~K \cite{Cao1998}.

\noindent\textbf{Optical pump}
For both tr-REXS and tr-RIXS experiments 100~fs pump pulses were generated at 620~meV (2~$\mathrm{\mu}$m) using an optical parametric amplifier. The pulses were polarized vertically in the $ab$-plane of the sample and were incident at 13$^{\circ}$ with respect to the sample surface. The choice of pump energy follows previous optical conductivity measurements \cite{Moon2006} and resonates between the upper and lower Hubbard bands. 

\noindent\textbf{The time-resolved resonant elastic X-ray scattering (tr-REXS) setup.} %
The tr-REXS experiment was performed at beamline 2 of the SPring-8 Angstrom Compact free electron LAser (SACLA) with a 30~Hz pulse repetition rate. We adopted a horizontal scattering geometry as seen in Fig.~\ref{Fig1}a and tuned the X-ray energy to the peak in the Ir $L_3$-edge resonance around 11.215~keV. A Multi-Port Charged Coupled Device (MPCCD) area detector was placed at $2\theta = 88.7^{\circ}$ to observe the magnetic Bragg peak $(-3, -2, 28)$. This geometry is chosen to optimize the X-ray resonant magnetic scattering cross section. We access the magnetic peak by rotating the sample around the vertical axis by $\phi = 12.8^{\circ}$ with the infrared and X-ray photons in an approximately co-linear geometry. The detector was read-out shot-by-shot and the signal was thresholded to suppress the background coming from X-ray fluorescence and electrical noise. The peak intensity was determined by binning the 2D MPCCD data into a 1D spectrum and fitting a Lorentzian lineshape with a uniform offset background. Each datapoint is the result of summing 1000-4000 shots. Previous characterization of the beamline found that the time resolution of this experiment was jitter-limited to approximately 300~fs.

The minimal model for the fitting is outlined in the main text (Equation \ref{equation}). This formula was convolved with a 100~fs Gaussian to account for the pump pulse width. The other major contribution to the effective time resolution was the x-ray pulse jitter of approximately 300~fs, because this is only an approximate value this was not included in the fit, rather this is taken as an upper limit on the decay time. Apart from this quantity, all parameters were varied to fit the data in the long time delay scans at 1.0, 2.7 and 13.8~mJ/cm$^2$ fluence in Fig.~\ref{Fig2}c and these fits were used to constrain $\tau_{\text{3D}}$ in fits of the short time delay data in Fig.~\ref{Fig2}b by interpolating the variation of $\tau_{\text{3D}}$ and $C$ as a function of fluence. In this way, equation \ref{equation} provides an accurate parametrization of the recovery dynamics at all fluences studied.

\noindent\textbf{The time-resolved resonant inelastic X-ray scattering (tr-RIXS) setup.}
The tr-RIXS experiment was performed at the X-ray Pump Probe instrument at the Linac Coherent Light Source (LCLS) with a 120~Hz repetition rate. We adopt a horizontal scattering plane, similar to the setup in the tr-REXS experiment. The $(\pi,0)$ and $(\pi,\pi)$ were measured at $(-3.5, -3.5,  24.1)$ and $(-4, -3, 23.9)$. Non-integer values of $L$ were chosen to keep the X-ray incident angle around $1^\circ$ as the RIXS spectrum is known to be essentially independent of $L$ \cite{Kim2012}. A Si $(333)$ monochromator produced a 50~meV incident energy bandpass. The RIXS spectrometer is conceptually similar to that used at Sector 27 at the Advanced Photon Source. Scattered photons from the sample are reflected from a segmented spherical Si$(8, 4, 4)$ analyzer in a near-backscattering configuration and detected by a Princeton CCD. The sample, the analyzer crystal, and the photodetector are placed on a Rowland circle with a radius of 1~m in the vertical plane. The total energy resolution of the tr-RIXS setup was $\sim70$~meV and the $\mathbi{Q}$ resolution was defined by the $\sim 6^\circ$ angular acceptance of the analyzer. RIXS spectra were collected in a stationary mode without moving the spectrometer and the pixel-to-energy conversion was performed using well-established methods. The CCD was read out every 1800 shots.  Jitter was the main contribution to the time resolution, which was on the order of 500~fs.

\section*{Acknowledgments}
We thank Robert Konik,  Weiguo Yin and Vittorio Cataudella for discussions.  Work performed at Brookhaven National Laboratory was supported by the US Department of Energy, Division of Materials Science, under Contract Numbers DE-SC00112704 and DE-AC02-98CH10886 and M.P.M.D.'s Early Career Award 1047478. X.L.\ acknowledges financial support from MOST (No.\ 2015CB921302) and CAS (Grant No: XDB07020200) of China. P.J.\ acknowledges support by Laboratory Directed Research and Development (LDRD) Program 12-007 (Complex Modeling). J.K., D.C.\ and A.S.\ were supported by the U.S.\ Department of Energy under Contract No.\ DE-AC02-06CH11357. S.W.\ acknowledges financial support from Spanish MINECO (Severo Ochoa grant SEV-2015-0522),  Ramon y Cajal program RYC-2013-14838, Marie Curie Career Integration Grant PCIG12-GA-2013-618487 and and Fundaci\'{o} Privada Cellex. J.L.\ is sponsored by the Science Alliance Joint Directed Research and Development Program at the University of Tennessee. Work in London was supported by the EPSRC. The magnetic Bragg peak measurements were performed at the BL3 of SACLA with the approval of the Japan Synchrotron Radiation Research Institute (JASRI) (Proposal No.\ 2014B8018). Use of the Linac Coherent Light Source (LCLS), SLAC National Accelerator Laboratory, is supported by the U.S. Department of Energy, Office of Science, Office of Basic Energy Sciences under Contract No.\ DE-AC02-76SF00515.

\section*{Author contributions}
J.P.H., X.L., M.P.M.D.\ and M.F. initiated and planned the project. M. P.M.D., Y.C., X.L., S.W., D.Z., R.M., V.T. X.M.C., J.V., D.C., J.K., A.H.S., P.J., R.A.-M., M.G., A.R., J.R., M.S., S.S., M.K., H.L., L.P., S.O., T.K., M.Y., Y.T., T.T., L.H., C.-L.C., D.F.M., M.F.\ and J.P.H. prepared for and performed the experiments. M.P.M.D., Y.C., X.L., S.W., M.F., D.F.M.\ and J.P.H. analyzed and interpreted the data. J.L., C.R.S.\ and B.J.K. prepared the samples. M.P.M.D.\ and Y.C.\ wrote the paper with contributions from X.L., S.W.,  D.F.M., M.F.\ and J.P.H.


\begin{thebibliography}{10}
\expandafter\ifx\csname url\endcsname\relax
  \def\url#1{\texttt{#1}}\fi
\expandafter\ifx\csname urlprefix\endcsname\relax\def\urlprefix{URL }\fi
\providecommand{\bibinfo}[2]{#2}
\providecommand{\eprint}[2][]{\url{#2}}

\bibitem{Scalapino2012}
\bibinfo{author}{Scalapino, D.~J.}
\newblock \bibinfo{title}{A common thread: The pairing interaction for
  unconventional superconductors}.
\newblock \emph{\bibinfo{journal}{Rev. Mod. Phys.}}
  \textbf{\bibinfo{volume}{84}}, \bibinfo{pages}{1383--1417}
  (\bibinfo{year}{2012}).

\bibitem{Kim2014}
\bibinfo{author}{Kim, Y.} \emph{et~al.}
\newblock \bibinfo{title}{Fermi arcs in a doped pseudospin-1/2 {Heisenberg}
  antiferromagnet}.
\newblock \emph{\bibinfo{journal}{Science}} \textbf{\bibinfo{volume}{345}},
  \bibinfo{pages}{187--190} (\bibinfo{year}{2014}).

\bibitem{Cao2014}
\bibinfo{author}{Cao, Y.} \emph{et~al.}
\newblock \bibinfo{title}{Hallmarks of the mott-metal crossover in the hole
  doped {$J= 1/2$ Mott} insulator {Sr$_2$IrO$_4$}}.
\newblock \emph{\bibinfo{journal}{arXiv preprint arXiv:1406.4978}}
  (\bibinfo{year}{2014}).

\bibitem{delaTorre2015}
\bibinfo{author}{de~la Torre, A.} \emph{et~al.}
\newblock \bibinfo{title}{Collapse of the mott gap and emergence of a nodal
  liquid in lightly doped {${\mathrm{Sr}}_{2}{\mathrm{IrO}}_{4}$}}.
\newblock \emph{\bibinfo{journal}{Phys. Rev. Lett.}}
  \textbf{\bibinfo{volume}{115}}, \bibinfo{pages}{176402}
  (\bibinfo{year}{2015}).

\bibitem{Fausti2001}
\bibinfo{author}{Fausti, D.} \emph{et~al.}
\newblock \bibinfo{title}{Light-induced superconductivity in a stripe-ordered
  cuprate}.
\newblock \emph{\bibinfo{journal}{Science}} \textbf{\bibinfo{volume}{331}},
  \bibinfo{pages}{189--191} (\bibinfo{year}{2011}).

\bibitem{Zhang2014}
\bibinfo{author}{Zhang, J.} \& \bibinfo{author}{Averitt, R.}
\newblock \bibinfo{title}{Dynamics and control in complex transition metal
  oxides}.
\newblock \emph{\bibinfo{journal}{Annual Review of Materials Research}}
  \textbf{\bibinfo{volume}{44}}, \bibinfo{pages}{19--43}
  (\bibinfo{year}{2014}).

\bibitem{Aoki2014}
\bibinfo{author}{Aoki, H.} \emph{et~al.}
\newblock \bibinfo{title}{Nonequilibrium dynamical mean-field theory and its
  applications}.
\newblock \emph{\bibinfo{journal}{Rev. Mod. Phys.}}
  \textbf{\bibinfo{volume}{86}}, \bibinfo{pages}{779--837}
  (\bibinfo{year}{2014}).

\bibitem{Ament2011}
\bibinfo{author}{Ament, L. J.~P.}, \bibinfo{author}{van Veenendaal, M.},
  \bibinfo{author}{Devereaux, T.~P.}, \bibinfo{author}{Hill, J.~P.} \&
  \bibinfo{author}{van~den Brink, J.}
\newblock \bibinfo{title}{Resonant inelastic x-ray scattering studies of
  elementary excitations}.
\newblock \emph{\bibinfo{journal}{Rev. Mod. Phys.}}
  \textbf{\bibinfo{volume}{83}}, \bibinfo{pages}{705--ï¿œ767}
  (\bibinfo{year}{2011}).

\bibitem{Kim2009}
\bibinfo{author}{Kim, B.~J.} \emph{et~al.}
\newblock \bibinfo{title}{Phase-sensitive observation of a spin-orbital mott
  state in {Sr$_2$IrO$_4$}}.
\newblock \emph{\bibinfo{journal}{Science}} \textbf{\bibinfo{volume}{323}},
  \bibinfo{pages}{1329--1332} (\bibinfo{year}{2009}).

\bibitem{Kim2012}
\bibinfo{author}{Kim, J.} \emph{et~al.}
\newblock \bibinfo{title}{Magnetic excitation spectra of
  {${\mathrm{Sr}}_{2}{\mathrm{IrO}}_{4}$} probed by resonant inelastic x-ray
  scattering: Establishing links to cuprate superconductors}.
\newblock \emph{\bibinfo{journal}{Phys. Rev. Lett.}}
  \textbf{\bibinfo{volume}{108}}, \bibinfo{pages}{177003}
  (\bibinfo{year}{2012}).

\bibitem{Lee2006}
\bibinfo{author}{Lee, P.~A.}, \bibinfo{author}{Nagaosa, N.} \&
  \bibinfo{author}{Wen, X.-G.}
\newblock \bibinfo{title}{Doping a {Mott} insulator: Physics of
  high-temperature superconductivity}.
\newblock \emph{\bibinfo{journal}{Rev. Mod. Phys.}}
  \textbf{\bibinfo{volume}{78}}, \bibinfo{pages}{17--85}
  (\bibinfo{year}{2006}).

\bibitem{Wang2011}
\bibinfo{author}{Wang, F.} \& \bibinfo{author}{Senthil, T.}
\newblock \bibinfo{title}{Twisted hubbard model for {Sr$_2$IrO$_4$}:
  {M}agnetism and possible high temperature superconductivity}.
\newblock \emph{\bibinfo{journal}{Phys. Rev. Lett.}}
  \textbf{\bibinfo{volume}{106}}, \bibinfo{pages}{136402}
  (\bibinfo{year}{2011}).

\bibitem{Yan2015}
\bibinfo{author}{Yan, Y.~J.} \emph{et~al.}
\newblock \bibinfo{title}{Electron-doped ${\mathrm{sr}}_{2}{\mathrm{iro}}_{4}$:
  An analogue of hole-doped cuprate superconductors demonstrated by scanning
  tunneling microscopy}.
\newblock \emph{\bibinfo{journal}{Phys. Rev. X}} \textbf{\bibinfo{volume}{5}},
  \bibinfo{pages}{041018} (\bibinfo{year}{2015}).

\bibitem{Cao1998}
\bibinfo{author}{Cao, G.}, \bibinfo{author}{Bolivar, J.},
  \bibinfo{author}{McCall, S.}, \bibinfo{author}{Crow, J.~E.} \&
  \bibinfo{author}{Guertin, R.~P.}
\newblock \bibinfo{title}{Weak ferromagnetism, metal-to-nonmetal transition,
  and negative differential resistivity in single-crystal
  {${\mathrm{Sr}}_{2}{\mathrm{IrO}}_{4}$}}.
\newblock \emph{\bibinfo{journal}{Phys. Rev. B}} \textbf{\bibinfo{volume}{57}},
  \bibinfo{pages}{R11039--R11042} (\bibinfo{year}{1998}).

\bibitem{Moon2006}
\bibinfo{author}{Moon, S.~J.} \emph{et~al.}
\newblock \bibinfo{title}{Electronic structures of layered perovskite
  {${\mathrm{Sr}}_{2}M{\mathrm{O}}_{4}$} ({$M=\mathrm{Ru}$, Rh, and Ir})}.
\newblock \emph{\bibinfo{journal}{Phys. Rev. B}} \textbf{\bibinfo{volume}{74}},
  \bibinfo{pages}{113104} (\bibinfo{year}{2006}).

\bibitem{Ehrke2011}
\bibinfo{author}{Ehrke, H.} \emph{et~al.}
\newblock \bibinfo{title}{Photoinduced melting of antiferromagnetic order in
  {${\mathrm{La}}_{0.5}{\mathrm{Sr}}_{1.5}{\mathrm{MnO}}_{4}$} measured using
  ultrafast resonant soft x-ray diffraction}.
\newblock \emph{\bibinfo{journal}{Phys. Rev. Lett.}}
  \textbf{\bibinfo{volume}{106}}, \bibinfo{pages}{217401}
  (\bibinfo{year}{2011}).

\bibitem{Zhou2012}
\bibinfo{author}{Zhou, S.} \emph{et~al.}
\newblock \bibinfo{title}{Glass-like recovery of antiferromagnetic spin
  ordering in a photo-excited manganite {Pr$_{0.7}$Ca$_{0.3}$MnO$_3$}}.
\newblock \emph{\bibinfo{journal}{Scientific reports}}
  \textbf{\bibinfo{volume}{4}} (\bibinfo{year}{2014}).
\newblock \bibinfo{note}{Doi:10.1038/srep04050}.

\bibitem{Chuang2013}
\bibinfo{author}{Chuang, Y.~D.} \emph{et~al.}
\newblock \bibinfo{title}{Real-time manifestation of strongly coupled spin and
  charge order parameters in stripe-ordered
  {${\mathrm{La}}_{1.75}{\mathrm{Sr}}_{0.25}{\mathrm{NiO}}_{4}$} nickelate
  crystals using time-resolved resonant x-ray diffraction}.
\newblock \emph{\bibinfo{journal}{Phys. Rev. Lett.}}
  \textbf{\bibinfo{volume}{110}}, \bibinfo{pages}{127404}
  (\bibinfo{year}{2013}).

\bibitem{Caviglia2013}
\bibinfo{author}{Caviglia, A.~D.} \emph{et~al.}
\newblock \bibinfo{title}{Photoinduced melting of magnetic order in the
  correlated electron insulator {NdNiO$_3$}}.
\newblock \emph{\bibinfo{journal}{Phys. Rev. B}} \textbf{\bibinfo{volume}{88}},
  \bibinfo{pages}{220401} (\bibinfo{year}{2013}).

\bibitem{lee2012phase}
\bibinfo{author}{Lee, W.-S.} \emph{et~al.}
\newblock \bibinfo{title}{Phase fluctuations and the absence of topological
  defects in a photo-excited charge-ordered nickelate}.
\newblock \emph{\bibinfo{journal}{Nature communications}}
  \textbf{\bibinfo{volume}{3}}, \bibinfo{pages}{838} (\bibinfo{year}{2012}).

\bibitem{Boeglin2010}
\bibinfo{author}{Boeglin, C.} \emph{et~al.}
\newblock \bibinfo{title}{Distinguishing the ultrafast dynamics of spin and
  orbital moments in solids}.
\newblock \emph{\bibinfo{journal}{Nature}} \textbf{\bibinfo{volume}{465}},
  \bibinfo{pages}{458--461} (\bibinfo{year}{2010}).

\bibitem{Kampfrath2011}
\bibinfo{author}{Kampfrath, T.} \emph{et~al.}
\newblock \bibinfo{title}{Coherent terahertz control of antiferromagnetic spin
  waves}.
\newblock \emph{\bibinfo{journal}{Nature Photonics}}
  \textbf{\bibinfo{volume}{5}}, \bibinfo{pages}{31--34} (\bibinfo{year}{2011}).

\bibitem{Malinowski2008}
\bibinfo{author}{Malinowski, G.} \emph{et~al.}
\newblock \bibinfo{title}{Control of speed and efficiency of ultrafast
  demagnetization by direct transfer of spin angular momentum}.
\newblock \emph{\bibinfo{journal}{Nature Physics}}
  \textbf{\bibinfo{volume}{4}}, \bibinfo{pages}{855--858}
  (\bibinfo{year}{2008}).

\bibitem{Dean2015}
\bibinfo{author}{Dean, M.~P.~M.}
\newblock \bibinfo{title}{Insights into the high temperature superconducting
  cuprates from resonant inelastic x-ray scattering}.
\newblock \emph{\bibinfo{journal}{Journal of Magnetism and Magnetic Materials}}
  \textbf{\bibinfo{volume}{376}}, \bibinfo{pages}{3 -- 13}
  (\bibinfo{year}{2015}).

\bibitem{Batignani2015}
\bibinfo{author}{Batignani, G.} \emph{et~al.}
\newblock \bibinfo{title}{Probing ultrafast photo-induced dynamics of the
  exchange energy in a heisenberg antiferromagnet}.
\newblock \emph{\bibinfo{journal}{Nature Photonics}}  (\bibinfo{year}{2015}).
\newblock \bibinfo{note}{Doi:10.1038/nphoton.2015.121}.

\bibitem{Ishii2011}
\bibinfo{author}{Ishii, K.} \emph{et~al.}
\newblock \bibinfo{title}{Momentum-resolved electronic excitations in the mott
  insulator {${\mathrm{Sr}}_{2}{\mathrm{IrO}}_{4}$} studied by resonant
  inelastic x-ray scattering}.
\newblock \emph{\bibinfo{journal}{Phys. Rev. B}} \textbf{\bibinfo{volume}{83}},
  \bibinfo{pages}{115121} (\bibinfo{year}{2011}).

\bibitem{Kim2014excitonic}
\bibinfo{author}{Kim, J.} \emph{et~al.}
\newblock \bibinfo{title}{Excitonic quasiparticles in a spin--orbit mott
  insulator}.
\newblock \emph{\bibinfo{journal}{Nature communications}}
  \textbf{\bibinfo{volume}{5}} (\bibinfo{year}{2014}).

\bibitem{Okamoto2010}
\bibinfo{author}{Okamoto, H.} \emph{et~al.}
\newblock \bibinfo{title}{Ultrafast charge dynamics in photoexcited
  {${\text{Nd}}_{2}{\text{CuO}}_{4}$} and {${\text{La}}_{2}{\text{CuO}}_{4}$}
  cuprate compounds investigated by femtosecond absorption spectroscopy}.
\newblock \emph{\bibinfo{journal}{Phys. Rev. B}} \textbf{\bibinfo{volume}{82}},
  \bibinfo{pages}{060513} (\bibinfo{year}{2010}).

\bibitem{Manousakis1991}
\bibinfo{author}{Manousakis, E.}
\newblock \bibinfo{title}{The spin-$1/2$ heisenberg antiferromagnet on a square
  lattice and its application to the cuprous oxides}.
\newblock \emph{\bibinfo{journal}{Rev. Mod. Phys.}}
  \textbf{\bibinfo{volume}{63}}, \bibinfo{pages}{1--62} (\bibinfo{year}{1991}).

\bibitem{Ronnow2001}
\bibinfo{author}{R\o{}nnow, H.~M.} \emph{et~al.}
\newblock \bibinfo{title}{Spin dynamics of the 2d spin $\frac{1}{2}$ quantum
  antiferromagnet copper deuteroformate tetradeuterate (cftd)}.
\newblock \emph{\bibinfo{journal}{Phys. Rev. Lett.}}
  \textbf{\bibinfo{volume}{87}}, \bibinfo{pages}{037202}
  (\bibinfo{year}{2001}).

\bibitem{Fujiyama2012}
\bibinfo{author}{Fujiyama, S.} \emph{et~al.}
\newblock \bibinfo{title}{Two-dimensional heisenberg behavior of
  {$J_{\text{eff}}= 1/2$} isospins in the paramagnetic state of the
  spin-orbital {Mott} insulator {Sr$_2$IrO$_4$}}.
\newblock \emph{\bibinfo{journal}{Phys. Rev. Lett.}}
  \textbf{\bibinfo{volume}{108}}, \bibinfo{pages}{247212}
  (\bibinfo{year}{2012}).

\bibitem{Vale2015}
\bibinfo{author}{Vale, J.~G.} \emph{et~al.}
\newblock \bibinfo{title}{Importance of {$XY$} anisotropy in
  {${\mathrm{Sr}}_{2}{\mathrm{IrO}}_{4}$} revealed by magnetic critical
  scattering experiments}.
\newblock \emph{\bibinfo{journal}{Phys. Rev. B}} \textbf{\bibinfo{volume}{92}},
  \bibinfo{pages}{020406} (\bibinfo{year}{2015}).

\bibitem{Serrao2013}
\bibinfo{author}{Rayan~Serrao, C.} \emph{et~al.}
\newblock \bibinfo{title}{Epitaxy-distorted spin-orbit mott insulator in
  {Sr$_{2}$IrO$_{4}$} thin films}.
\newblock \emph{\bibinfo{journal}{Phys. Rev. B}} \textbf{\bibinfo{volume}{87}},
  \bibinfo{pages}{085121} (\bibinfo{year}{2013}).

\end{thebibliography}
\end{document}